\documentstyle[11pt,moriond,epsfig]{article}

\def\Journal#1#2#3#4{{#1} {\bf #2}, #3 (#4)}

\def\PLB{{\em Phys. Lett.}  B}
\def\PRL{\em Phys. Rev. Lett.}
\def\PRD{{\em Phys. Rev.} D}

\def\be{\begin{equation}}
\def\ee{\end{equation}}
\def\bea{\begin{eqnarray}}
\def\eea{\end{eqnarray}}

\begin{document}
\begin{flushright}
CLNS 99/1623
\end{flushright}
\vspace*{4cm}
\title{CKM PHYSICS FROM $B$ DECAYS USING THE CLEO EXPERIMENT}

\author{ {\sc Andreas Warburton}\\(representing the CLEO Collaboration)\\
\vspace*{0.5cm}}
\address{Floyd R. Newman Laboratory of Nuclear Studies,\\ Cornell University,\\
Ithaca, New York  14853,\\ USA}

\maketitle\abstracts{ We report on studies of three types of $B$-meson
decay that can contribute to an understanding of fundamental
intergenerational quark mixing, charge-conjugation--parity violation,
and long-distance quantum chromodynamics.  Specifically, we discuss a
selection of analyses related to the Cabibbo-Kobayashi-Maskawa (CKM)
parameters $V_{cb}$, $V_{ub}$, $\alpha$, and $\gamma$, and the
nonperturbative heavy quark effective theory quantities $\lambda_1$
and $\overline{\Lambda}$.  We first describe an examination of the
first and second moments of the hadronic-recoil mass and
charged-lepton energy spectra in inclusive $b\to c\,\ell\,\nu$ decays.
We also report on the reconstruction, using similar experimental
techniques, of the CKM-suppressed decay $B\to\rho\,\ell\,\nu$ and the
extraction of its branching fraction, ${\cal
B}(B^0\to\rho^-\,\ell^+\,\nu) = (2.57 \pm 0.29\;^{+0.33}_{-0.46}\pm
0.41) \times 10^{-4}$, as well as the value $|V_{ub}| = (3.25 \pm
0.14\;^{+0.21}_{-0.29} \pm 0.55) \times 10^{-3}$, where the
uncertainties are statistical, systematic, and due to model
dependence, respectively.  Finally, we present results on rare
two-body charmless hadronic $B\to K\pi$, $\pi\pi$, and $KK$ decays and
comment briefly on their implications to the geometry of the CKM
unitarity triangle, including a bound on $\gamma$.}

\vspace*{0.5in}
\begin{center}
June, 1999
\end{center}

\vspace*{1.0in}
\begin{center}
{\it Invited talk presented at the 1999 Electroweak Interactions and
Unified Theories \\ session of the XXXIV$^{\rm th}$ Rencontres de
Moriond, \\ Arc 1800, Savoie, France}
\end{center}

\newpage
\section{Introduction}


The flavour-dependent strengths of the weak interactions of quarks can
be expressed in terms of the Cabibbo-Kobayashi-Maskawa (CKM) mixing
matrix~\cite{ckm} $V_{\rm CKM}$, which, by convention, rotates the
$\pm 1/3$-charged quark mass states into their weak eigenstates.
Under the constraints that there be three quark flavour generations
and that the CKM matrix be unitary, this mixing can be expressed in
terms of four fundamental constants of nature, including a parameter
$\eta$ that allows for charge-conjugation--parity ($CP$) violation:
$e.g.$~\cite{ckm},
\begin{equation}
\label{eqn:ckm}
V_{\rm CKM}
=
\left(\begin{array}{ccc}
        V_{ud} & V_{us} & V_{ub} \\
        V_{cd} & V_{cs} & V_{cb} \\
        V_{td} & V_{ts} & V_{tb}
\end{array}\right)
\simeq
\left(\begin{array}{ccc}
        1 - \lambda^2/2 & \lambda & A\lambda^3(\rho - i\eta(1-\lambda^2/2)) \\
        -\lambda & 1-\lambda^2/2-i\eta A^2\lambda^4 & A\lambda^2(1+i\eta\lambda^2) \\
        A\lambda^3(1-\rho-i\eta) & -A\lambda^2 & 1
\end{array}\right).
\end{equation}
The pursuit of measurements to overconstrain the CKM matrix by
extracting its parameters from several observables constitutes a
significant fraction of contemporary experimental programmes; however,
the determination of these parameters in the presence of the
confounding effects of long-distance quantum chromodynamics (QCD) and
non-tree-level processes requires considerable theoretical input.

We briefly present a selection of studies conducted at the Cornell
Electron Storage Ring (CESR) and with direct implications for CKM
physics.  The $4\pi$ solenoidal CLEO detector~\cite{cleo}, comprising
tracking chambers, a CsI electromagnetic calorimeter, and muon
systems, is situated at the CESR $e^+\,e^-$ interaction region, where
$B\overline{B}$ meson pairs are produced near threshold by decays of
the $\sim$10.58~GeV/$c^2$ $\Upsilon(4S)$ bottomonium resonance.
Continuum production, $e^+\, e^- \to q\bar{q}$ ($q \in \{u, d, s,
c\}$), with approximately three times the effective cross section of
$\Upsilon(4S)$ production ($\sigma_{\Upsilon(4S)} \simeq 1.07$~nb),
forms the principal source of background for the decays discussed here
and is statistically subtracted using data collected $\sim$60
MeV/$c^2$ below the $\Upsilon(4S)$ resonance.  Unless noted otherwise,
the data sample used for the results in this paper corresponds to a
time-integrated luminosity of $\int\!\!{\cal L}dt \simeq
3.1$~fb$^{-1}$ ($\sim$3.3$\times 10^6$ $B\overline{B}$ candidates)
collected near the $\Upsilon(4S)$ and $\sim$1.6~fb$^{-1}$ taken off
resonance.

\section{Moments Analysis of Inclusive $b\to c\,\ell\,\nu$ Decays:
A Path to $V_{cb}$}

Inclusive rate measurements of $b\to c\,\ell\,\nu$ processes can
furnish information on $V_{cb}$.  Heavy quark effective theory (HQET)
and an operator product expansion (OPE) have been used to compute
model-independent inclusive semileptonic $B$-meson decay rates as a
series in powers of $\alpha_s$ and $\Lambda_{\rm QCD} / m_b$ in the
$m_b \to \infty$ limit~\cite{hqet}.  These expansions contain
corrections in the form of hadronic matrix elements denoted by
$\lambda_1$ and $\lambda_2$, which physically represent the squared
average momentum of the $b$ quark inside its meson and the energy of
the hyperfine interaction of the $b$ quark's spin with that of the
light degrees of freedom, respectively.  The latter quantity is known
to be $\lambda_2 \simeq 0.12$~(GeV/$c^2$)$^2$ from the $B^* - B$ mass
splitting.  A third parameter, $\overline{\Lambda}$, exists in the
HQET expansions to relate the $b$-quark and $B$-meson masses:
$\overline{\Lambda} \equiv m_B - m_b + \frac{\lambda_1 +
3\lambda_2}{2m_b} + {\cal O}(\Lambda^3_{\rm QCD} / m^2_b)$.

In order to extract $V_{cb}$ from rate studies and HQET, the
quantities $\lambda_1$ and $\overline{\Lambda}$ must first be
determined experimentally using measurements of other observables
computed using the OPE.  Expressions have been calculated to ${\cal
O}(1/m^2_b)$ for the first and second moments observables of both the
hadronic recoil mass~\cite{had_mom1,had_mom2,had_mom3} ($m_{X_c}$) and
the charged-lepton energy~\cite{lept_mom} ($E_\ell$) in $B\to
X_c\,\ell\,\nu$ decays.  We report here on recent preliminary CLEO
results of the extracted regions of the $\lambda_1 -
\overline{\Lambda}$ plane from measurements of these four moments.

\subsection{Hadronic Mass Moments}
\label{sec:hadmom}

\begin{figure}
\begin{center}
\rotatebox{-90}{\epsfig{figure=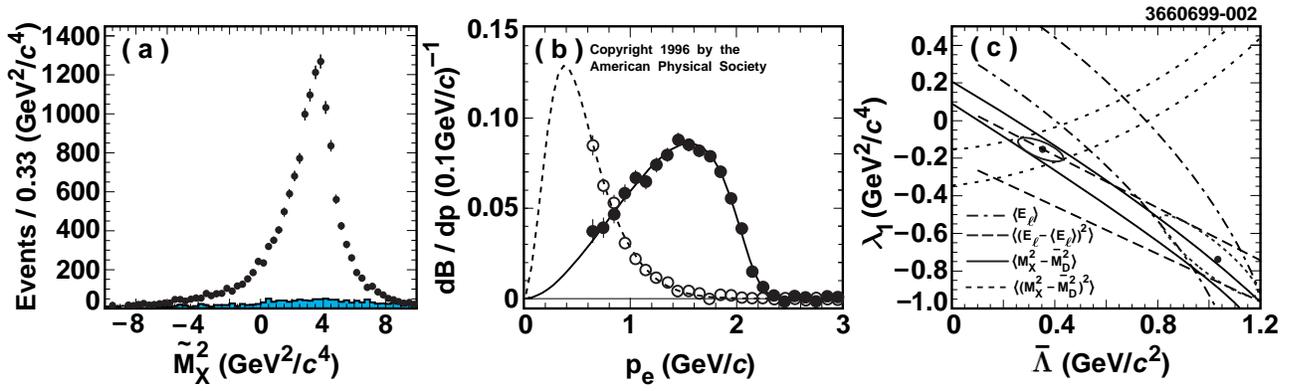,height=6.75in}}
\end{center}
\caption
{\label{fig:moments}
(a) The square of the measured mean mass of the hadronic
recoil system.  The points represent data taken near the
$\Upsilon(4S)$ mass resonance; the shaded histogram denotes scaled
off-resonance data.  (b) The primary electron momentum distribution
(filled circles) of candidate $B\to X\, e\,\nu$ decays from a
$\sim$2.06~fb$^{-1}$ data sample~\protect\cite{apsfig}.  The open
circles are for secondary leptons from $b\to c$ decays and the curves
are fits to the modified ISGW model~\protect\cite{isgw}.  (c) The
extracted allowed ($1\sigma$) regions in the $\lambda_1 -
\overline{\Lambda}$ HQET parameter plane based on preliminary
experimental results for the first and second lepton-energy and
hadronic-mass moments.  Uncertainties are correlated between the
bands.}
\end{figure}

To measure the hadronic mass moments in $B\to X_c\,\ell\,\nu$ decays,
we first selected events containing a single charged-lepton candidate
with momentum $1.5 < p_\ell < 2.5$~GeV/$c$.  The kinematics of the
neutrino were inferred~\cite{nurecon} using the relative hermeticity
of the CLEO detector and the well-known $e^+\,e^-$ beam energy.  The
square of the hadronic recoil mass was then computed using the
kinematics of the $\ell$ and $\nu$ candidates: $m_{X_c}^2 = m_B^2 +
m_{\ell\nu}^2 - 2E_B E_{\ell\nu} +
2\left|\vec{p}_B\right|\left|\vec{p}_{\ell\nu}\right|
\cos\theta_{\ell\nu{\rm ,}B}$, where $\theta_{\ell\nu{\rm ,}B}$ is the
angle in the lab frame between the flight directions of the $B$
candidate and the lepton system.  In practice, we used the expression
$\widetilde{m}_{X_c}^2 \equiv m_B^2 + m_{\ell\nu}^2 - 2E_B
E_{\ell\nu}$, because of the unknown direction of $\vec{p}_B$ and its
relatively small magnitude ($\sim$300~MeV/$c$).  The measured
$\widetilde{m}_{X_c}^2$ distribution, which is shown in
Fig.~\ref{fig:moments}(a), is dominated ($\sim$96\%) by $b\to
c\,\ell\,\nu$ processes.  The remaining contributions, $\sim$3\% from
$b\to c\to s\,\ell\,\nu$ secondary decays and charmonium leptons and
$\sim$1\% from $b\to u\,\ell\,\nu$ processes, were subtracted using
Monte Carlo calculations.  After correcting for bias arising from the
difference between $\widetilde{m}_{X_c}^2$ and $m_{X_c}^2$ and from
the asymmetry in the neutrino momentum resolution, we find the first
and second hadronic mass moments to be $\langle m^2_{X_c} -
\overline{m}^2_D \rangle = 0.286 \pm 0.023 \pm 0.080$~(GeV/$c^2$)$^2$
and $\langle (m^2_{X_c} - \overline{m}^2_D)^2 \rangle = 0.911 \pm
0.066 \pm 0.309$~(GeV/$c^2$)$^4$, respectively, where the first
uncertainties are statistical and the second systematic.  The moments
are calculated with respect to the spin-averaged charm meson
mass, $\overline{m}_D = 1.975$~GeV/$c^2$.

Solving for the nonperturbative HQET parameters in the theoretical
moments expressions~\cite{had_mom1,had_mom3} yields the following
solutions: $\overline{\Lambda} = 0.33 \pm 0.02 \,[{\rm stat}] \pm 0.08
\,[{\rm syst}] \ {\rm GeV/}c^2$ and $\lambda_1 = -0.13 \pm 0.01
\,[{\rm stat}] \pm 0.06 \,[{\rm syst}] \ ({\rm GeV/}c^2)^2$.  The
preliminary measured allowed bands in the $\lambda_1 -
\overline{\Lambda}$ plane are shown in Fig.~\ref{fig:moments}(c)).

\subsection{Charged-Lepton Energy Moments}

The charged-lepton energy ($E_\ell$) moments were determined from the
primary electron momentum distribution~\cite{apsfig} depicted in
Fig.~\ref{fig:moments}(b).  Candidate electrons with momenta greater
than 0.6~GeV/$c$ were identified as primary or secondary by examining
charge and angular correlations with an additional high-momentum
($>$1.4~GeV/$c$) charged-lepton candidate.  We corrected the spectrum
in Fig.~\ref{fig:moments}(b) for radiative and resolution effects not
included in the OPE prediction, for the boost of the $B$ meson, and
for the extrapolation to momenta below 0.6~GeV/$c$.  The resultant
preliminary first and second moments are $\langle E_\ell \rangle =
1.36\pm 0.01\pm 0.02$~GeV and $\langle \left( E_\ell - \langle E_\ell
\rangle \right)^2\rangle = 0.190\pm 0.004\pm 0.005$~(GeV)$^2$; the
corresponding $\lambda_1 - \overline{\Lambda}$ bands determined from a
theoretical expansion~\cite{lept_mom} are illustrated in
Fig.~\ref{fig:moments}(c).

The apparent discrepancy between the preliminary hadron-mass and
charged-lepton-energy moments results in Fig.~\ref{fig:moments}(c)
needs to be understood before a reliable extraction of $V_{cb}$ can be
achieved using the heavy quark expansion.  Sizeable theoretical
uncertainties have been estimated from calculations of ${\cal
O}(1/m^3_b)$ contributions~\cite{had_mom2,had_mom3,bauer} and could
account for much of the inconsistency between the $\lambda_1 -
\overline{\Lambda}$ bands for the hadron and lepton moments results.
In fact, the use of the second hadron-mass moment to determine
$\overline{\Lambda}$ and $\lambda_1$, as was done in
Sec.~\ref{sec:hadmom}, has been discouraged due to
radiative-correction and ${\cal O}(1/m_b^3)$
effects~\cite{had_mom3,bauer}.  Moreover, the sensitivity of the
$\overline{\Lambda}$ and $\lambda_1$ parameters to the model-dependent
extrapolation into the region $E_\ell \leq 0.6$~GeV may compromise the
reliability of the lepton-energy moment measurements~\cite{zoltan}.
Finally, the linear combinations of $\overline{\Lambda}$ and
$\lambda_1$ constrained in the hadron-mass and charged-lepton moments
studies are effectively the same, rendering a simultaneous solution of
these HQET parameters unfeasible.  An additional observable, {\it
e.g.}, the first moment in the photon energy spectrum in $B\to
X_s\,\gamma$ (with a different linear combination of
$\overline{\Lambda}$ and $\lambda_1$), will help complete the
picture~\cite{kapustin_ligeti,had_mom3,bauer}.

\section{The CKM-Suppressed Decay $B\to\rho\,\ell\,\nu$ and a New
Measurement of $|V_{ub}|$}

The decay $B\to\rho\,\ell\,\nu$ is sensitive to the suppressed element
$V_{ub}$ of the CKM matrix.  Experimental study of $b\to u$ processes
is challenged by the relatively low decay rates and the significant
backgrounds from $b\to c$ sources.  We therefore measure a partial
rate for $B\to\rho\,\ell\,\nu$ decay in the charged-lepton endpoint
region, $E_\ell > 2.3$~GeV, in which there is negligible phase space
for $b\to c\,\ell\,\nu$ channels.  The extrapolation to the total rate
from the partial rate in the endpoint region introduces model
dependence, since a knowledge of the shape of the hadronic form-factor
distribution is necessary.  Further significant model dependence
enters in the extraction of $|V_{ub}|$ from the total rate, as
estimates of the normalization $\Gamma / \left| V_{ub} \right|^2$ are
needed.  Studies of $q^2$, the square of the mass of the virtual $W$
boson in the decay, can help to reduce the $|V_{ub}|$ theory
dependence by constraining the form-factor models.  We report on new
results for the branching fraction, $|V_{ub}|$ extraction, and $q^2$
distribution in $B\to\rho\,\ell\,\nu$ decays~\cite{lange}.

The analysis technique used a simultaneous binned maximum-likelihood
fit in several variables: three bins of $E_\ell$ in the ranges
[1.7$-$2.0], [2.0$-$2.3], and [2.3$-$2.7]~GeV; five $b\to
u\,\ell\,\nu$ signal modes, namely the hadronic final states $\rho$
($\pi^\pm\,\pi^0$ or $\pi^+\,\pi^-$), $\omega$
($\pi^+\,\pi^-\,\pi^0$), $\pi^\pm$, and $\pi^0$; the kinematic
variables $\Delta E \equiv \left( E_\rho + E_\ell + \left|
\vec{p}_{\rm miss}\right|\right) - E_{\rm beam}$ and $m_\rho$; and
background contributions from continuum, cross feed, fake leptons, and
$b\to u$ (other than $\rho$, $\omega$, and $\pi$) and $b\to c$
sources.  The quantities $\vec{p}_{\rm miss} \equiv -\sum\vec{p}_i
\approx \vec{p}_\nu$ (where $i$ runs over reconstructed charged tracks
and CsI energy clusters in the event) and $E_{\rm beam}$ represent the
candidate neutrino momentum~\cite{nurecon} and the beam energy,
respectively.  Several event-shape criteria~\cite{lange} were used to
suppress continuum events, which constituted the principal background.
\begin{figure}
\begin{center}
\epsfig{figure=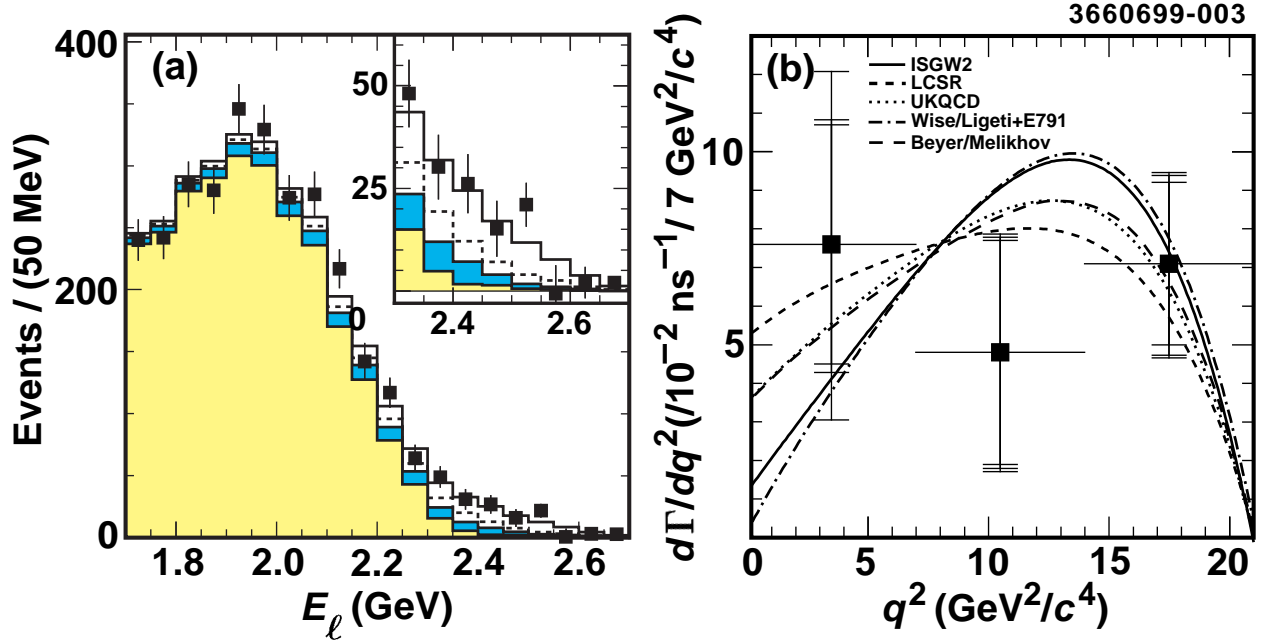,width=6.75in}
\end{center}
\caption{(a) Fit projection of the kinematic variable $E_\ell$, with
$|\Delta E| < 500$~MeV and $|m_{\pi\pi} - m_\rho| < 150$~MeV/$c^2$.
The points are on-resonance data after continuum subtraction and the
histogram is the fit projection with contributions from signal
(uppermost unshaded), $b\to c$ (light shading), non-signal $b\to u$
(dark shading), and signal cross feed (unshaded region below dashed
line).  The inset shows the endpoint region. (b) Comparison of the
measured $\Delta\Gamma$ distribution (points) with predictions from
the form-factor models~\protect\cite{ff_models} after extrapolation to
the full $E_\ell$ range.}
\label{fig:rholnu}
\end{figure}

Fig.~\ref{fig:rholnu}(a) shows the fit projection of the
charged-lepton energy $E_\ell$, whereas Fig.~\ref{fig:rholnu}(b)
compares the $q^2$-dependent $B\to\rho\,\ell\,\nu$ partial width
distributions for 5 form-factor models~\cite{ff_models} with the
measured widths in three $q^2$ bins (extrapolated to all values of
$E_\ell$):
\begin{eqnarray}
\Delta\Gamma(q^2 < 7 \; ({\rm GeV})^2/c^4) & = &
(7.6\pm 3.0\,[{\rm stat}]\,^{+0.9}_{-1.2}\,[{\rm syst}]\pm 3.0\,[{\rm model}])
  \times 10^{-2} \; {\rm ns}^{-1} \nonumber\\
\Delta\Gamma(7 \leq q^2 < 14 \; ({\rm GeV})^2/c^4) & = &
(4.8\pm 2.9\,[{\rm stat}]\,^{+0.7}_{-0.8}\,[{\rm syst}]\pm 0.7\,[{\rm model}])
  \times 10^{-2} \; {\rm ns}^{-1} \\
\Delta\Gamma(q^2 \geq 14 \; ({\rm GeV})^2/c^4) & = &
(7.1\pm 2.1\,[{\rm stat}]\,^{+0.9}_{-1.1}\,[{\rm syst}]\pm 0.6\,[{\rm model}])
  \times 10^{-2} \; {\rm ns}^{-1}.\nonumber
\end{eqnarray}
Fig.~\ref{fig:rholnu}(b) suggests that more data and, in particular,
experimental studies in the $E_\ell < 2.3$~GeV region will be needed
if the form-factor models are to be confronted.  We also use the 5
form-factor models~\cite{ff_models} to determine the mean
$B\to\rho\,\ell\,\nu$ branching fraction and $|V_{ub}|$, where we have
taken a quadratic sum of half the spread in the individual model
results and a 15\% error for $\Gamma/|V_{ub}|^2$ as the theoretical
uncertainty.  After including a previous CLEO
measurement~\cite{nurecon} in our averages, we get the results
\begin{eqnarray}
{\cal B}(B^0\to\rho^-\,\ell^+\,\nu) & = &
(2.57\pm 0.29\,[{\rm stat}]\,^{+0.33}_{-0.46}\,[{\rm syst}]\pm
	0.41\,[{\rm model}]) \times 10^{-4} \\
|V_{ub}| & = &
(3.25\pm 0.14\,[{\rm stat}]\,^{+0.21}_{-0.29}\,[{\rm syst}]\pm
	0.55\,[{\rm model}]) \times 10^{-3}. \nonumber
\end{eqnarray}

\section{Charmless Hadronic $B\to K\pi$, $\pi\pi$, and $KK$ Decays:
Clues about Penguins, $\alpha$, and $\gamma$}

\begin{figure}
\begin{center}
\epsfig{figure=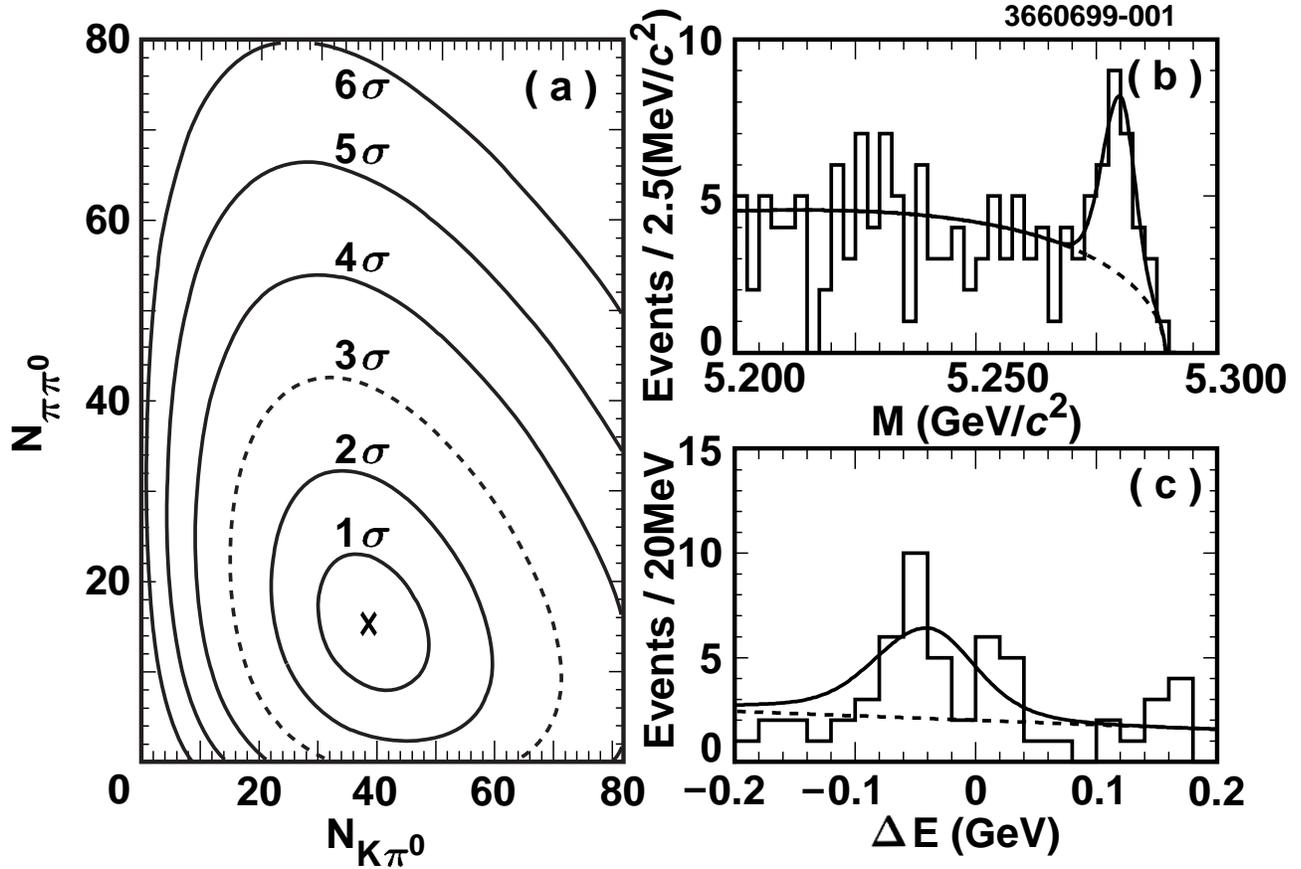,width=6.75in}
\end{center}
\caption{(a) Yield contours for $B^\pm\to\pi^\pm\,\pi^0$ and $B^\pm\to
K^\pm\,\pi^0$ modes in the maximum-likelihood fit.  (b) Projection of
beam-constrained mass $M$ for $B^\pm\to K^\pm\,\pi^0$ candidates.  (c)
Projection of $\Delta E$ for $B^\pm\to K^\pm\,\pi^0$ candidates.  Note
that there exists an expected $-42$~MeV offset because pion mass
hypotheses were used in the reconstruction.}
\label{fig:Kpi0}
\end{figure}

The imposition of unitarity on Eq.~\ref{eqn:ckm} can be described in
terms of a triangle with internal angles $\alpha$, $\beta$, and
$\gamma$, providing a geometric description of $CP$ violation.  Rare
charmless hadronic $B$ decays are a fertile source of information on
$\alpha$ and $\gamma$ and can probe non-Standard-Model and
non-tree-level processes.  CLEO has recently reported new results on
charmless hadronic $B$ decays~\cite{gao_frank}.  Here we briefly
summarize the measurements of $B\to K\pi$, $\pi\pi$, and $KK$ decays
in a sample of $\sim$5.8 million $B\overline{B}$ pairs.  Tracks, which
are required to satisfy several quality criteria, are identified as
pions or kaons based on their specific ionization characteristics.
Using pairs of tracks and $\pi^0$ candidates, we calculate the
beam-constrained mass, $M\equiv \sqrt{E^2_{\rm beam} - p^2_B}$, where
$p_B$ is the $B$ momentum.  We also define $\Delta E \equiv \sum E_j -
E_{\rm beam}$, where $j$ indexes the daughters.  Several event-shape
variables are used to separate signal events from continuum, which is
the main source of background.  The signal yields are determined by
computing unbinned maximum-likelihood fits in all these variables;
refer to Fig.~\ref{fig:Kpi0} for some example projections.  The
branching-fraction results are summarized in
Table~\ref{tab:br_fractions}.
\begin{table}
\caption{Preliminary CLEO efficiency, yield, significance, and
branching-fraction results.\label{tab:br_fractions}}
\vspace{0.4cm}
\begin{center}
\begin{tabular}{|c||c|c|c|c|}
\hline
Mode & Efficiency (\%) & Yield & Significance & ${\cal B} \times 10^5$
		\\\hline\hline
$K^\pm\,\pi^\mp$  & $53\pm 5$ & $43.1\,^{+9.0}_{-8.2}$ & $>6\sigma$ &
	$1.4\pm 0.3\,[{\rm stat}]\pm 0.2\,[{\rm syst}]$ \\
$K^\pm\,\pi^0$    & $42\pm 4$ & $38.1\,^{+9.7}_{-8.7}$ & $>6\sigma$ &
	$1.5\pm 0.4\,[{\rm stat}]\pm 0.3\,[{\rm syst}]$ \\
$K^0\,\pi^\pm$    & $15\pm 2$ & $12.3\,^{+4.7}_{-3.9}$ & $>5\sigma$ &
	$1.4\pm 0.5\,[{\rm stat}]\pm 0.2\,[{\rm syst}]$ \\\hline
$\pi^\pm\,\pi^\mp$& $53\pm 5$ & $11.5\,^{+6.3}_{-5.2}$ & $<3\sigma$ &
	$<0.84$ (90\% C.L.) \\
$\pi^\pm\,\pi^0$  & $42\pm 4$ & $14.9\,^{+8.1}_{-6.9}$ & $<3\sigma$ &
	$<1.6\phantom{0}$ (90\% C.L.) \\\hline
$K^\pm\,K^\mp$    & $53\pm 5$ & $\phantom{0}0.0\,^{+1.6}_{-0.0}$ &  &
	$<0.23$ (90\% C.L.) \\
$K^\pm\,K^0$      & $15\pm 2$ & $\phantom{0}1.8\,^{+2.6}_{-1.4}$ &  &
	$<0.93$ (90\% C.L.) \\\hline
\end{tabular}
\end{center}
\end{table}

The mode $B\to\pi^+\,\pi^-$, dominated by $b\to u$ tree processes, is
attractive because of its potential in providing $\alpha$ via $B^0 -
\overline{B^0}$ time-dependent mixing; however, penguin pollution and
rescattering effects introduce theoretical uncertainties that, in
order to be understood, require measurements of other processes such
as $B_s\to K^+\,K^-$ decays or other $B\to \pi\,\pi$
modes~\cite{pirjol_babar}.  Moreover, the low upper limit on ${\cal
B}(B\to\pi^+\,\pi^-)$ raises the possibility of a significant strong
phase, which has hitherto been taken to be zero, between isospin
amplitudes~\cite{gao_frank}.

The use of the principally gluonic-penguin $B\to K\pi$ modes for the
extraction of information on $\gamma$, currently the most uncertain
CKM parameter, has been the subject of recent intense theoretical
interest.  Several $\gamma$-bounding proposals involving $K\pi$
branching-fraction ratios exist~\cite{gamma_theory,neubert_rosner},
but are blighted by one or more of electroweak penguins, final-state
interactions, or an incomplete knowledge of the relative spectator and
penguin decay amplitudes.  We have used the method of Neubert and
Rosner~\cite{neubert_rosner}, with our preliminary measurement of the
ratio $R_* \equiv {\cal B}(B^\pm\to K^0\,\pi^\pm) / 2{\cal B}(B^\pm\to
K^\pm\,\pi^0) = 0.47\pm 0.24$, to determine a bound $\cos\gamma \leq
0.33$ at the 90\% confidence level~\cite{gao_frank}.  Preliminary
analyses of $B\to K\pi$, $\pi\pi$, and $KK$ candidates using the full
CLEO data set of $\sim$10 million $B\overline{B}$ pairs will be
available later this year and are expected to augment the results
presented here.

\section*{Acknowledgments}
My colleagues in the CLEO collaboration and the staff at CESR made
these results possible.  I thank V\'{e}ronique Boisvert, David Lange,
Michael Luke, and Frank W\"{u}rthwein for useful discussions.  I am
grateful to the Natural Sciences and Engineering Research Council of
Canada, the Training and Mobility of Researchers Programme of the
European Union, and Cornell University for their support.

\end{document}